\documentclass{myaa}

\usepackage[varg]{txfonts}

\usepackage{natbib}
\usepackage{graphicx}
\usepackage{epstopdf}
\usepackage{amssymb}
\usepackage{float}

\begin{document}

\title{Theoretical model of HD 163296 presently forming in-situ planets and comparison with the models of AS 209, HL Tau, and TW Hya}

\author{Dimitris M. Christodoulou\inst{1,2}  
\and 
Demosthenes Kazanas\inst{3}
}

\institute{
Lowell Center for Space Science and Technology, University of Massachusetts Lowell, Lowell, MA, 01854, USA.\\
\and
Dept. of Mathematical Sciences, Univ. of Massachusetts Lowell, 
Lowell, MA, 01854, USA. \\ E-mail: dimitris\_christodoulou@uml.edu\\
\and
NASA/GSFC, Laboratory for High-Energy Astrophysics, Code 663, Greenbelt, MD 20771, USA. \\ E-mail: demos.kazanas@nasa.gov \\
}


\def\gsim{\mathrel{\raise.5ex\hbox{$>$}\mkern-14mu
                \lower0.6ex\hbox{$\sim$}}}

\def\lsim{\mathrel{\raise.3ex\hbox{$<$}\mkern-14mu
               \lower0.6ex\hbox{$\sim$}}}

\abstract{
We fit an isothermal oscillatory density model to the disk of HD 163296 in which planets have presumably already formed and they are orbiting at least within the four observed dark gaps. This 156 AU large axisymmetric disk shows various physical properties comparable to those of AS 209, HL Tau, and TW Hya that we have modeled previously; but  it compares best to AS 209. The disks of HD 163296 and AS 209 are comparable in size and they share similar values of the power-law index $k\approx 0$ (a radial density profile of the form $\rho(R)\propto R^{-1}$), the rotational parameter $\beta_0$ (to within a factor of 3); a relatively small inner core radius (although this parameter for HD 163296 is exceptionally small, $R_1\simeq 0.15$ AU, presumably due to unresolved planets in the inner 50 AU); the scale length $R_0$ and the Jeans gravitational frequency $\Omega_J$ (to within factors of 1.4); the equation of state ($c_0^2/\rho_0$) and the central density $\rho_0$ (to within factors of 2); and the core angular velocity $\Omega_0$ (to within a factor of 4.5). In the end, we compare all six nebular disks that we have modeled so far. 
}

\authorrunning{ }
\titlerunning{Planet formation in the disk of HD 163296}

\maketitle

\section{Introduction}\label{intro}

In previous work \citep{chr19a,chr19b,chr19c,chr19d}, we presented isothermal models of the solar nebula and four ALMA/DSHARP observed protostellar disks capable of forming protoplanets long before their protostars will actually be formed by gas accretion/dispersal processes. This new ``bottom-up'' planet formation scenario is currently observed in real time by the latest high-resolution ($\sim$1-5~AU) observations of many protostellar disks by the ALMA telescope \citep{alm15,and16,rua17,lee17,lee18,mac18,ave18,cla18,kep18,guz18,ise18,zha18,dul18,fav18,har18,hua18,per18,kud18,lon18,pin18,vdm19}.   

The ALMA/DSHARP observations show many circular protostellar disks with annular dark gaps presumably carved out by protoplanets that have already been formed at a time long before accretion/dispersal processes will dissipate these gaseous disks. Up until now, few disks show asymmetries and spiral arms, signs of developing instabilities \citep{per18,hua18,vdm19,vil19}. Motivated by the DSHARP observations of annular gaps, we have produced theoretical models of the disks of AS 209 (seven gaps), HL Tau (seven gaps), TW Hya (5 gaps), and RU Lup (4 gaps) \citep{chr19b,chr19c,chr19d}. In this work, we apply the same theoretical model to the observed disk of HD 163296, a young protostellar system observed by ALMA/DSHARP \citep{hua18}. 

HD 163296 turns out to be difficult to model. It shows only four widely-spaced dark gaps over a radial extent of 156 pc, and it is evident that its inner 50 pc are not adequately resolved by the ALMA observations \citep{hua18}. The analytic (intrinsic) and numerical (oscillatory) solutions of the isothermal Lane-Emden equation \citep{lan69,emd07} with differential rotation, and the resulting model of the midplane of the gaseous disk have been described in detail in \cite{chr19a} for the solar nebula. Here, we apply in \S~\ref{models2} the same model to the four dark gaps of HD 163296. In \S~\ref{disc}, we summarize our modeling results and we compare the inferred physical parameters of HD 163296 against those found previously for AS 209, HL Tau, and TW Hya.

\section{Physical Models of the HD 163296 Protostellar Disk}\label{models2}

\subsection{Numerical Setup}

The numerical integrations that produce oscillatory density profiles were performed with the \textsc{Matlab} {\tt ode15s} integrator \citep{sha97,sha99} and the optimization used the Nelder-Mead simplex algorithm as implemented by \cite{lag98}. This method (\textsc{Matlab} routine {\tt fminsearch}) does not use any numerical or analytical gradients in its search procedure which makes it extremely stable numerically, albeit somewhat slow. The boundary conditions for the oscillatory density profiles are, as usual, $\tau(0)=1$ and $[d\tau/dx](0)=0$, where $\tau$ and $x$ are the dimensionless values of the density and the radius, respectively.

\begin{figure}
\begin{center}
    \leavevmode
      \includegraphics[trim=0.2 0.2cm 0.2 0.2cm, clip, angle=0, width=10 cm]{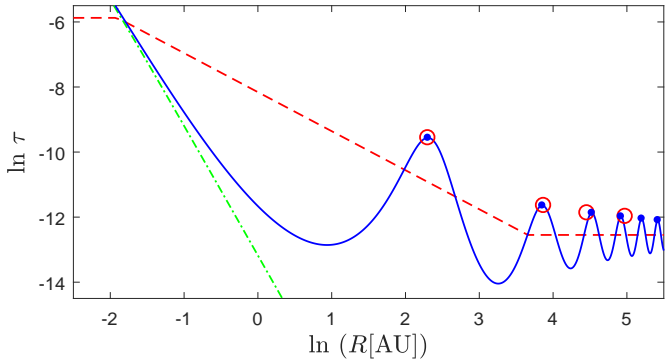}
      \caption{Equilibrium density profile for the midplane of the HD 163296 disk that has already formed at least four annular dark gaps (presumably protoplanets) \citep{hua18}. The best-fit parameters are $k=-0.2$, $\beta_0=0.0530$, $R_1=0.1492$~AU, and $R_2=38.79$~AU. The radial scale length of the disk is $R_0=0.01312$~AU. The Cauchy solution (solid line) has been fitted to the dark gaps of HD 163296 (Table~\ref{table1}) so that its density maxima (dots) correspond to the observed orbits of the protoplanets (open circles). The density maximum corresponding to the location of the first maximum was scaled to a distance of 10~AU of the D10 gap. The mean relative error of the fit is 6.3\%, most of it coming from the two outer gaps (Table~\ref{table1}). The intrinsic analytical solution (dashed line) and the nonrotating analytical solution (dash-dotted line) are also shown for reference. Notice the exceptionally small inner core ($\ln R_1 = -1.90$) of the model. 
\label{fig1}}
  \end{center}
\end{figure}

\begin{table*}
\caption{Radii of dark gaps in AS 209, HL Tau, TW Hya, and HD 163296 \citep[from Table 1 of][]{hua18}}
\label{table1}
\begin{tabular}{ll|ll|ll|ll}
\hline
Gap   & AS 209     & Gap    &  HL Tau & Gap  &  TW Hya & Gap & HD 163296 \\
Name & $R~(AU)$ & Name  &  $R~(AU)$ & Name & $R~(AU)$ &Name  &  $R~(AU)$ \\
\hline
D9   &   08.69   &   D14   &   13.9 & D1 & 1         & D10   & 10  \\
D24  &  23.84  &   D34   &   33.9 & D26 &  25.62 & D48   & 48  \\
D35  &  35.04  &   D44   &   44 & D32 & 31.5       & D86   & 86.4  \\
D61  &  60.8    &   D53   &   53 & D42 &  41.64    & D145  & 145  \\
D90  &  89.9    &   D67   &   67.4 & D48 & 48       &    &   \\
D105 &  105.5  &   D77   &   77.4 & &                  &    &   \\
D137 &  137     &   D96   &   96 & &                     &    &   \\
\hline
\end{tabular}
\end{table*}

\begin{table*}
\caption{Comparison of the protostellar disk model of HD 163296 against AS 209, HL Tau, and TW Hya}
\label{table2}
\begin{tabular}{llllll}
\hline
Property & Property & AS 209 & HL Tau & TW Hya & HD 163296 \\
Name     & Symbol (Unit) & Best-Fit Model & Best-Fit Model & Best-Fit Model&Best-Fit Model \\
\hline
Density power-law index & $k$                                          &   $0.0$  	     & $0.0$   & $-0.2$ & $-0.2$ \\
Rotational parameter & $\beta_0$                                &    0.0165	       &  0.00562 &  0.00401 &  0.0530 \\
Inner core radius & $R_1$ (AU)                              &   6.555  	       &  52.04  &  28.67 &  0.1492\\
Outer flat-density radius & $R_2$ (AU)                              &   68.96        	   &  90.55  & $\cdots$ & 38.79 \\
Scale length & $R_0$ (AU)    &   0.01835  &  0.009813  & 0.004100  & 0.01312 \\
Equation of state & $c_0^2/\rho_0$ (${\rm cm}^5 {\rm ~g}^{-1} {\rm ~s}^{-2}$) & $6.32\times 10^{16}$ & $1.81\times 10^{16}$ &  $3.15\times 10^{15}$  & $3.23\times 10^{16}$ \\
Minimum core density$^a$ & $\rho_0$ (g~cm$^{-3}$)         &    $5.62\times 10^{-9}$   			&   $1.97\times 10^{-8}$ &  $1.13\times 10^{-7}$ & $1.10\times 10^{-8}$  \\
Isothermal sound speed$^a$ & $c_0$ (m~s$^{-1}$) & 188 & 188 & 188 & 188 \\
Jeans gravitational frequency & $\Omega_J$ (rad~s$^{-1}$)    &    $4.9\times 10^{-8}$ & $9.1\times 10^{-8}$  &  $2.2\times 10^{-7}$ & $6.8\times 10^{-8}$ \\
Core angular velocity & $\Omega_0$ (rad~s$^{-1}$)    &    $8.0\times 10^{-10}$ 	& $5.1\times 10^{-10}$  & $8.7\times 10^{-10}$  & $3.6\times 10^{-9}$ \\
Core rotation period & $P_0$ (yr)                                 &    249 	   			&  390  & 228 & 55.3 \\
Maximum disk size & $R_{\rm max}$ (AU)                &    144 	   			&   102 & 52 &  156\\
\hline
\end{tabular}
$^a$Calculated for $T=10$~K and $\overline{\mu} = 2.34$
\end{table*} 

\begin{figure}
\begin{center}
    \leavevmode
      \includegraphics[trim=0.2 1.5cm 0.2 1.5cm, clip, angle=0, width=10 cm]{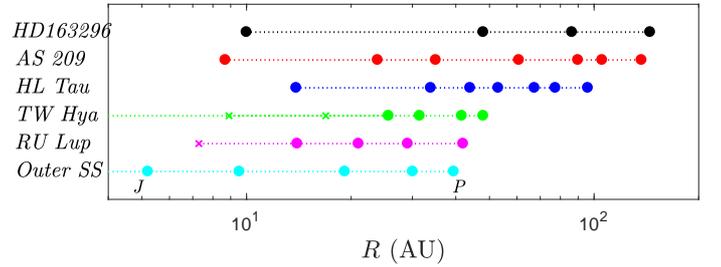}
      \caption{Schematic diagram of the ALMA-observed dark gaps (dots) that we have modeled so far. Key: {\it SS}:Solar System, {\it J}:Jupiter, {\it P}:Pluto. The crosses represent the empty density peaks in which no dark gaps have been observed yet. The dark gaps of HD 163296 are depicted by black dots. The disk of this protostellar system is comparable in size to the disk of AS 209, and the calculated physical properties of the best-fit model indicate that the HD 163296 model is more similar to AS 209 than any other model (Table~\ref{table2}).
\label{fig_disks_2}}
  \end{center}
\end{figure}

\subsection{Best-Fit model of HD 163296}\label{model1}

The radii of the four dark gaps observed in HD 163296 are shown in Table~\ref{table1}. In Fig.~\ref{fig1}, we show the best optimized fit to these four gaps. In the models, we could not produce any acceptable result with only three free parameters ($k$, $\beta_0$, and $R_1$). We had to also introduce a flat-density region starting at radius $R_2$, and it turns out that three of four gaps fall within this outer intrinsically flat region. The mean relative error of the fit is 6.3\% and it comes mostly from the two outermost gaps D86 and D145 (Table~\ref{table1}).

The physical properties of the best-fit HD 163296 model are listed in Table~\ref{table2} along with the best-fit models of AS 209, HL Tau, and TW Hya. Although the best-fit model of HD 163296 shares similar parameters with all of these other protostellar disks, it is apparent that it is most similar to the best-fit model of AS 209. A detailed comparison between the disks of HD 163296 and AS 209 is made in \S~\ref{disc} below.

\section{Summary and Discussion}\label{disc}

In \S~\ref{models2}, we presented the best-fit isothermal differentially-rotating protostellar model of HD 163296 observed by ALMA/DSHARP \citep{hua18}. This model shows four dark gaps in a large disk (Table~\ref{table1}), and it is widely believed that protoplanets have already formed and carved out these gaps in the observed axisymmetric disk. The best-fit model is depicted in Figure~\ref{fig1} and a comparison of its physical properties versus AS 209, HL Tau, and TW Hya is shown in Table~\ref{table2}. The physical properties of HD 163296 are much closer to those of AS 209 than any of the other disks listed in Table~\ref{table2}.

In Fig.~\ref{fig_disks_2}, we show a schematic diagram of dark gaps (dots) in the ALMA-observed disks that we have modeled up until now, and we have also included our outer solar system ({\it Outer SS}). There is no physics to be deduced from this figure (as opposed to Table~\ref{table2}), this layout is only a relative comparison of the arrangements of dark gaps in the depicted protostellar systems. In conjunction with the physical properties listed in Table~\ref{table2}, the picture that emerges about HD 163296 is the following:
\begin{enumerate}
\item[1.] In general, one can identify various isolated properties of HD 163296 that are similar to any other of these other disks.
\item[2.] However HD 163296 is most similar to AS 209 in its physical properties (Table~\ref{table2}) and in its layout of dark gaps (Fig.~\ref{fig_disks_2}), although it exhibits fewer gaps than AS 209. In particular, we note the following physical properties that are comparable between these two disks:
\item[2a.] The power-law index of HD 163296 is $k=-0.2$, close to zero as in AS 209. This implies a radial density profile of approximately $\rho(R)\propto R^{-1}$ for both disks.
\item[2b.] The inner core radius of HD 163296 has the smallest radius found during all of our modeling efforts. The inner radius is merely $R_1\simeq 0.15$ AU, as opposed to the relatively small inner radius of $\simeq 6.6$ AU for AS 209 and $\simeq 0.82$ AU for the solar nebula.
\item[2c.] The rotational parameter of HD 163296, $\beta_0 = 0.0530$, is so small that it guarantees the stability of the young disk against nonaxisymmetric dynamical instabilities \citep{chr95}.
\item[2d.] The scale length $R_0$ and the Jeans gravitational frequency $\Omega_J$ of the two disks are very much comparable to within factors of merely 1.4.
\item[2e.] The equation of state ($c_0^2/\rho_0$) and the central density $\rho_0$ of the two disks are comparable to within factors of about 2.
\item[2f.] The only fundamental parameter that is slightly larger in HD 163296 is the core's angular velocity $\Omega_0$; it is larger by a factor of 4.5 than in AS 209. We note however that the cores of both systems are not adequately resolved by the current ALMA/DSHARP observations \citep{hua18}.
\end{enumerate}
From these comparisons, it is evident that the protostellar disk of HD 163296 is quite similar in structure and physical properties to the disk of AS 209 (Table~\ref{table2}) and quite dissimilar to our solar nebula \citep{chr19a}.

\end{document}